\documentclass[doublecol]{epl2}

\usepackage{amsmath}

\usepackage{pdfpages}

\title{Stochastic single-molecule dynamics of synaptic membrane protein domains}
\shorttitle{Stochastic dynamics of synaptic membrane protein domains} 

\author{Osman Kahraman \and Yiwei Li \and Christoph A. Haselwandter}
\shortauthor{Osman Kahraman \etal}

\institute{Department of Physics \& Astronomy and Molecular and Computational Biology Program, \\Department of Biological Sciences, University of Southern California, Los Angeles, CA 90089, USA}

\pacs{87.16.-b}{Subcellular structure and processes}
\pacs{87.19.lp}{Pattern formation: activity and anatomic}
\pacs{87.10.Mn}{Stochastic modeling}

\abstract{
Motivated by single-molecule experiments on synaptic membrane protein domains, we use a stochastic lattice model to study protein reaction and diffusion processes in crowded membranes. We find that the stochastic reaction-diffusion dynamics of synaptic proteins provide a simple physical mechanism for collective fluctuations in synaptic domains, the molecular turnover observed at synaptic domains, key features of the single-molecule trajectories observed for synaptic proteins, and spatially inhomogeneous protein lifetimes at the cell membrane. Our results suggest that central aspects of the single-molecule and collective dynamics observed for membrane protein domains can be understood in terms of stochastic reaction-diffusion processes at the cell membrane.
}

\begin{document}

\maketitle

\section{Introduction}
A variety of essential biological functions of cell membranes rely on the organization of membrane proteins into membrane protein domains \cite{lang10,simons2010,rao2014,recouvreux2016}. Superresolution light microscopy \cite{choquet2003,huang2010,kusumi2014}
of membrane protein domains has shown that molecular diffusion can yield rapid stochastic turnover of individual membrane proteins, with complicated diffusion trajectories arising from molecular crowding and interactions between different protein species. A biologically important example of membrane protein domains is provided by synaptic domains \cite{ziv2014,salvatico2015},
which are crucial for signal transmission across chemical synapses. Synaptic domains are crowded with synaptic receptor and scaffold molecules, and mediate synaptic signaling via transient binding of synaptic receptors to neurotransmitter molecules released from the presynaptic terminal. The strength of the transmitted signal depends on the number of receptors localized in synaptic domains \cite{citri2008,specht2008}, and regulation of the receptor number in synaptic domains is one mechanism for postsynaptic plasticity~\cite{carroll2001,shepherd2007,kneussel2014}.

Synaptic domains of a well-defined characteristic size can persist over months or even longer periods of time \cite{trachtenberg2002, grutzendler2002}. However, receptor \cite{choquet2003,triller2008,triller2005} as well as scaffold \cite{okabe1999,gray2006,calamai2009} molecules have been observed \cite{choquet2013,kneussel2014}
to turn over rapidly, with individual molecules leaving and entering synaptic domains on typical timescales as short as seconds. Experiments \cite{kirsch1995,meier2000,meier2001,borgdorff2002,dahan2003,hanus2006,ehrensperger2007,calamai2009,haselwandter2011} and theoretical modeling \cite{haselwandter2011,haselwandter2015} have shown that the reaction and diffusion properties of synaptic receptors and their
associated scaffold molecules are sufficient for the spontaneous formation of synaptic domains, and that self-assembly of synaptic domains of the observed characteristic size can be understood in terms of a reaction-diffusion (Turing) instability \cite{turing1952}. Experiments \cite{ribrault2011,choquet2013} and theoretical modeling \cite{shouval2005,holcman2006,sekimoto2009,burlakov2012,czondor2012} also suggest that synaptic domains undergo collective fluctuations that may affect synaptic signaling. It is largely unknown how the rapid
stochastic dynamics of individual synaptic receptors and scaffolds \cite{meier2001,borgdorff2002,dahan2003,hanus2006,triller2005,specht2008,triller2008}
relate \cite{ribrault2011,choquet2013,ziv2014,salvatico2015} to the observed collective properties of synaptic domains.

In this letter we show that key features of the observed stochastic dynamics of synaptic domains can be understood in terms of a simple stochastic lattice model of receptor and scaffold reaction-diffusion processes at the membrane
\cite{haselwandter2011,haselwandter2015}, and thereby demonstrate emergence of synaptic domains in the presence of rapid stochastic turnover of individual molecules. Our stochastic lattice model yields excellent agreement with mean-field models \cite{haselwandter2011,haselwandter2015,satulovsky1996,mckane2004,lugo2008,fanelli2010,fanelli2013} of nonlinear diffusion in crowded membranes, but we find substantial discrepancies between mean-field and stochastic models for the reaction dynamics at synaptic domains. Kinetic Monte Carlo (KMC) simulations of our stochastic lattice model yield, in agreement with previous experiments and mean-field calculations
\cite{kirsch1995,meier2000,meier2001,borgdorff2002,dahan2003,hanus2006,ehrensperger2007,calamai2009,haselwandter2011,haselwandter2015}, spontaneous formation of synaptic domains, and demonstrate that the molecular noise induced by the underlying reaction and diffusion dynamics of synaptic
receptors and scaffolds can produce collective fluctuations in synaptic domains \cite{ribrault2011,choquet2013}. 
We show that, based on the reaction and diffusion properties of synaptic receptors and scaffolds suggested by previous experiments and mean-field calculations 
\cite{kirsch1995,meier2000,meier2001,borgdorff2002,dahan2003,hanus2006,ehrensperger2007,calamai2009,haselwandter2011,haselwandter2015},
our stochastic lattice model can yield the molecular turnover observed at synaptic domains \cite{choquet2003,specht2008,calamai2009}, predicts single-molecule trajectories consistent with experimental observations
\cite{choquet2003,choquet2013,meier2001,borgdorff2002,dahan2003,hanus2006,specht2008,triller2005,triller2008}, and provides a simple physical mechanism for spatially inhomogeneous receptor and scaffold lifetimes at the membrane \cite{czondor2012,blanpied2002,earnshaw2006}.
Thus, our stochastic lattice model links the molecular noise inherent in receptor-scaffold reaction-diffusion dynamics to collective fluctuations in synaptic domains and allows prediction of the stochastic dynamics of individual synaptic receptors and scaffolds, which cannot be achieved via existing mean-field models \cite{haselwandter2011,haselwandter2015}. While we focus here on synaptic domains as a model system, our main results are of broad applicability \cite{lang10,simons2010,rao2014,recouvreux2016,choquet2003} to the stochastic single-molecule dynamics of membrane protein domains.

\section{Reaction-diffusion dynamics} 
Membrane protein domains are characterized
\cite{lang10,simons2010,rao2014,recouvreux2016,choquet2003,specht2008,triller2005,triller2008,ribrault2011,choquet2013,ziv2014,salvatico2015}
by low protein copy numbers ($\approx10$--1000) while also providing highly crowded environments for reaction-diffusion processes to occur. We employ
\cite{satulovsky1996,mckane2004,lugo2008,
butler2009,butler2011,haselwandter2011,haselwandter2015,fanelli2010,fanelli2013,erban2009,cao2014} a stochastic lattice model of synaptic domains in which we divide the cell membrane into equal-sized patches (lattice sites) with reaction processes only occurring between receptors ($R$) and scaffolds ($S$) occupying the same membrane patch. For the sake of conceptual and computational simplicity, we focus here on the most straightforward scenario of a 1D system of length $L$ with periodic boundary conditions and $K$ patches of size $a=L/K$, and allow receptors and scaffolds to hop randomly to nearest-neighbor patches with hopping rates $1/\tau_{\alpha}$, where $\alpha=r,s$ for receptors and
scaffolds, respectively. We find that, consistent with previous work on stochastic reaction-diffusion models in population biology \cite{lugo2008}, this 1D formulation of the stochastic reaction-diffusion dynamics at synaptic domains \cite{haselwandter2011,haselwandter2015} already captures the basic phenomenology of the observed fluctuations at synaptic domains. Indeed, the 2D formulation of our model \cite{haselwandter2011,haselwandter2015} shows similar stochastic dynamics of synaptic domains as the 1D formulation we focus on here \cite{sm}. While not essential for capturing the basic phenomenology of fluctuations at synaptic domains, a 2D formulation would be required to make more detailed and quantitative comparisons with experimental results, which necessarily pertain to 2D systems.

Cell membranes provide highly crowded and heterogeneous molecular
environments, which can strongly affect protein reaction kinetics and give rise to anomalous diffusion of membrane proteins \cite{metzler2016,jeon2016}.
Based on previous work \cite{haselwandter2011,haselwandter2015,satulovsky1996,mckane2004,lugo2008,fanelli2010,fanelli2013} on reaction and diffusion processes in crowded environments, we use here a phenomenological model of crowding and assume that the rates of reaction and diffusion processes locally increasing the receptor or scaffold number are $\propto \left(1- N^r_i - N^s_i\right)$ for each site $i$, where $N^{\alpha}_i/\epsilon$ are the occupation numbers of receptors and scaffolds at site $i$ so that $0 \leq N^r_i + N^s_i \leq 1$ and each membrane patch can accommodate up to $1/\epsilon$ receptors or scaffolds. The physically relevant values of the normalization constant $\epsilon$ are coupled to the patch size $a$ and the size of the molecules under consideration, with $\epsilon$
decreasing with increasing $a$ and decreasing molecule size. We employ identical normalization constants $\epsilon$ for receptors and scaffolds, but distinct $\epsilon$ could be used for receptors and scaffolds to provide a more detailed model of molecule number in synaptic
domains. Our stochastic lattice model is defined mathematically by its master equation (ME) \cite{VanKampen2011, Gardiner2009},
\begin{equation}
\frac{\partial P}{\partial t}=\sum_{{\bf m}} \!\,\bigl[W({\bf N}-{\bf m};{\bf m}) P({\bf N}-{\bf m},t) - W({\bf N};{\bf m})P({\bf N},t)\bigr] \,,
\label{eq:master}
\end{equation}
where $P(\mathbf{N}, t)$ is the probability of the lattice occupancy $\mathbf{N}=\{\mathbf{N}^\alpha\}$ at time $t$ with $\mathbf{N}^\alpha= \left(N^\alpha_1, N^\alpha_2,\dots, N^\alpha_K\right)$, $W(\mathbf{N}; \mathbf{m})$ is the transition rate from $\mathbf{N}$ to $\mathbf{N}+\mathbf{m}$, and $\mathbf{m}$ is the array of jumps in receptor and scaffold lattice occupancies.

\begin{largetable}
\caption{Contributions to $\mathcal{R}_i^{(l)}$ in eq.~(\ref{eq:react}) and $F^{\alpha}(r,s)$ in eqs.~(\ref{eq:mfield1}) and~(\ref{eq:mfield2}) \cite{haselwandter2011,haselwandter2015}. Chemical reactions are expressed in terms of $R$ and $S$, which denote receptors and scaffolds at the cell membrane, $R_b$ and $S_b$, which stand for receptors and scaffolds in the bulk of the cell (cytoplasm), and the unspecified bulk molecule $M_b$, which models removal of a receptor or scaffold molecule from the membrane via some mechanism that involves a temporary increase in the local crowding of the cell membrane. We use the convention that $k_l$ denotes the rate of removal from/insertion into the membrane per molecule, and all $k_l$ are therefore scaled by $1/\epsilon$. We employ identical reaction kinetics and values of the dimensionless rate constants as in ref.~\cite{haselwandter2011}, which correspond to \textit{model C} in ref.~\cite{haselwandter2015} and are consistent with experiments on glycine receptors and gephyrin scaffolds \cite{specht2008,choquet2003,triller2008,triller2005}, with the time units set by the rate of receptor endocytosis $k_1=b=1/750$~$\text{s}^{-1}$. In particular, we use the parameter values $(m_1,m_2,\beta,\mu)=b(0.4,10,0.5,0.7)$ and $(\bar{r},\bar{s})=(0.05,0.05)$ \cite{haselwandter2011,haselwandter2015}, with the rightmost column in the table showing the connection between the notation used here and in refs.~\cite{haselwandter2011,haselwandter2015}.
(See refs.~\cite{haselwandter2011,haselwandter2015} for further details.)}
\label{tab1}
\begin{center}
\begin{tabular}{cccc}
\hline
Chemical reactions & Stochastic lattice model, $\mathcal{R}_i^{(l)}$ & Mean-field model, $F^{\alpha}(r,s)$ & Rates \\ \hline \hline
$R\xrightarrow{k_1} R_b$           & $\frac{k_1}{\epsilon} N^r_i$           & $-k_1 r$       & $k_1=b$\\
$R_b\xrightarrow{k_2} R$         & $\frac{k_2}{\epsilon}\left(1-N^r_i-N^s_i\right)$    & $k_2(1-r-s)$   & $k_2=m_1 \frac{\bar{r}}{1-\bar{r}-\bar{s}}$\\
$M_b+R\xrightarrow{k_3} M_b+R_b$ & $\frac{k_3}{\epsilon}\left(1-N^r_i-N^s_i\right) N^r_i$ & $-k_3(1-r-s)r$ & $k_3=\frac{m_1\bar{r}+m_2\bar{s}}{\bar{r}(1-\bar{r}-\bar{s})}$\\
$R_b+S\xrightarrow{k_4} R+S$     & $\frac{k_4}{\epsilon}\left(1-N^r_i-N^s_i\right) N^s_i$ & $k_4(1-r-s)s$  & $k_4=b\frac{\bar{r}}{\bar{s}}\frac{1}{1-\bar{r}-\bar{s}}$\\
$R_b+R+S\xrightarrow{k_5} 2R+S$  & $\frac{k_5}{\epsilon}\left(1-N^r_i-N^s_i\right) N^r_i N^s_i$ & $k_5(1-r-s)rs$ & $k_5=\frac{m_2}{\bar{r}}\frac{1}{1-\bar{r}-\bar{s}}$ \\ \hline 
$S\xrightarrow{k_6} S_b$           & $\frac{k_6}{\epsilon} N^s_i$           & $-k_6 s$       & $k_6=\beta$ \\
$S_b\xrightarrow{k_7} S$         & $\frac{k_7}{\epsilon}\left(1-N^r_i-N^s_i\right)$    & $k_7(1-r-s)$   & $k_7=\beta \frac{\bar{s}}{1-\bar{r}-\bar{s}}$\\
$M_b+S\xrightarrow{k_8} M_b+S_b$ & $\frac{k_8}{\epsilon} \left(1-N^r_i-N^s_i\right) N^s_i$ & $-k_8(1-r-s)s$ & $k_8=\frac{\mu}{1-\bar{r}-\bar{s}}$\\
$S_b+2S\xrightarrow{k_9} 3S$     & $\frac{k_9}{2! \epsilon} \left(1-N^r_i-N^s_i\right) N^s_i \left(N^s_i-\epsilon\right)$ & $\frac{k_9}{2!} (1-r-s) s^2$ & $k_9=\frac{\mu}{\bar{s}}\frac{2}{1-\bar{r}-\bar{s}}$ \\ \hline
\end{tabular}
\end{center}
\end{largetable}

The transition rate $W=W_\text{react}+W_\text{diff}$ in eq.~(\ref{eq:master}), where $W_\text{react}$ and $W_\text{diff}$ denote contributions to $W$ due to receptor and scaffold reaction and diffusion processes at the membrane. We have $W_\text{diff}=W_\text{diff}^r+W_\text{diff}^s$ with $W_\text{diff}^\alpha
=W_\text{diff}^{(1;\alpha)}+W_\text{diff}^{(2;\alpha)}$, in which $W_\text{diff}^{(1,2;\alpha)}$ are the transition rates for hopping from
site $i$ to sites $i\pm1$:
\begin{align} \label{eq:diff}
W_\text{diff}^{(1,2;\alpha)}({\bf N};{\bf m}) &= \frac{1}{2\tau_{\alpha} \epsilon} \sum_i N^\alpha_i (1-N^r_{i\pm1}-N^s_{i\pm1}) \nonumber 
\\ &\times \delta(m_i+\epsilon)\delta(m_{i\pm1}-\epsilon) \prod_{k\neq i,i\pm1} \delta(m_k) \, ,
\end{align}
where $\delta(x)$ is the Dirac $\delta$ function, and the factor of $1/\epsilon$
arises because we use the convention that $1/\tau_{\alpha}$ is the hopping
rate per molecule.

For the contributions to $W$ due to reactions we have $W_\text{react}=\sum_l W_\text{react}^{(l)}$, where each $W_\text{react}^{(l)}$
with $l=1,2,\dots$ corresponds to a particular reaction among receptors or scaffolds. The $W_\text{react}^{(l)}$ take the general form
\begin{align} \label{eq:react}
 W_\text{react}^{(l)}({\bf N};{\bf m}) = \sum_i \mathcal{R}_i^{(l)} \delta(m_i\pm \epsilon) \prod_{k\neq i} \delta(m_k) \, ,
\end{align}
where $\mathcal{R}_i^{(l)}$ encapsulates the properties of the specific reaction under consideration. For concreteness, we focus here on synaptic domains formed by glycine receptors and gephyrin scaffolds \cite{ribrault2011,choquet2013,ziv2014,salvatico2015}.
As explained in detail elsewhere \cite{haselwandter2011,haselwandter2015},
the experimental phenomenology of glycine receptors and gephyrin scaffolds
\cite{specht2008,choquet2003,triller2008,triller2005} suggests the reaction kinetics and values of the dimensionless rate constants summarized in table~\ref{tab1}, which we also used for the calculations described here. Key experimental features of the reaction-diffusion dynamics of glycine receptors and gephyrin for self-assembly of synaptic domains \cite{kirsch1995,meier2000,meier2001,borgdorff2002,dahan2003,hanus2006,ehrensperger2007,calamai2009,haselwandter2011} are \cite{specht2008,choquet2003,triller2008,triller2005,haselwandter2011,haselwandter2015} that glycine receptors diffuse rapidly compared to gephyrin, and that gephyrin can transiently bind glycine receptors as well as other gephyrin molecules.
While the reaction schemes in table~\ref{tab1} include the reactions between
glycine receptors and gephyrin suggested by experimental observations \cite{specht2008,choquet2003,triller2008,triller2005}, only a few of these reactions \cite{haselwandter2011,haselwandter2015}, such as trimerization of gephyrin \cite{calamai2009}, are essential for self-assembly of synaptic domains. We fixed the time units in our model by adjusting the rate of receptor endocytosis within the range of values estimated previously \cite{haselwandter2011,haselwandter2015} from experiments, which correspond to characteristic time scales ranging
from seconds to hours \cite{specht2008,choquet2003,triller2008}, to $k_1=1/750$~$\text{s}^{-1}$
so that our model reproduces the scaffold recovery time measured in fluorescence recovery after photobleaching (FRAP) experiments \cite{choquet2003,specht2008, calamai2009} (see below).

Molecular crowding and the reaction dynamics at synaptic domains 
\cite{choquet2003,triller2005,specht2008,triller2008,ribrault2011,choquet2013,ziv2014,salvatico2015}
make eq.~(\ref{eq:master}) with eqs.~(\ref{eq:diff}) and~(\ref{eq:react})
highly nonlinear \cite{haselwandter2011,haselwandter2015}. Direct solution of eq.~(\ref{eq:master}) with eqs.~(\ref{eq:diff}) and~(\ref{eq:react})
is therefore only practical for special cases. KMC simulations \cite{gillespie1976, gillespie1977,gillespie2013} provide a numerical procedure for circumventing these mathematical challenges. We implemented KMC simulations of eq.~(\ref{eq:master}) using Gillespie's ``direct'' method \cite{gillespie1976} and, unless indicated otherwise, employed random initial conditions of $\mathbf{N}$ satisfying $0 \leq N^r_i + N^s_i \leq 1$ for all $i$. We set $a\approx80$~nm and $\epsilon=1/100$ so that \cite{erban2009,cao2014} the membrane patch size was smaller than the expected typical size of synaptic domains \cite{kirsch1995,meier2000,meier2001,borgdorff2002,dahan2003,hanus2006,ehrensperger2007,calamai2009,haselwandter2011,haselwandter2015}
but large enough to accommodate multiple receptors and scaffolds, with size $\approx5$--$10$~nm for glycine receptors and gephyrin \cite{kim2006,du2015}.
To check for robustness we repeated our simulations for $\epsilon=1/200$--1/20 and $a=50$--100~nm, as well as for different $L$, and obtained similar results.

The mean-field equations describing receptor and scaffold reaction-diffusion dynamics can be derived \cite{VanKampen2011,haselwandter2015} from eq.~(\ref{eq:master})
with eqs.~(\ref{eq:diff}) and~(\ref{eq:react}), and are given by \cite{haselwandter2011}
\begin{eqnarray}
 \frac{\partial r}{\partial t} &=& F^r(r,s) +\nu_r \left[ (1-s)\nabla^2 r + r \nabla^2 s \right]\,, \label{eq:mfield1} \\
 \frac{\partial s}{\partial t} &=& F^s(r,s) +\nu_s \left[ (1-r)\nabla^2 s + s \nabla^2 r \right]\,,
 \label{eq:mfield2}
\end{eqnarray}
with all parameters determined directly by eq.~(\ref{eq:master}) with eqs.~(\ref{eq:diff}) and (\ref{eq:react}), where $r(x,t)$ and $s(x,t)$ are the deterministic continuum fields associated with $\mathbf{N}^{\alpha}$, the receptor and scaffold diffusion coefficients $\nu_{\alpha} = a^2/\left(2 \tau_{\alpha}\right)$, and the polynomials $F^{\alpha}(r,s)$ describe the mean-field reaction dynamics \cite{gillespie1976, gillespie1977,gillespie2013}. Table \ref{tab1} summarizes the additive contributions to $F^{\alpha}(r,s)$ implied by the standard formalism of chemical dynamics \cite{mckane2004,lugo2008,gillespie1976,gillespie1977,gillespie2013}
for the reaction kinetics considered here \cite{haselwandter2011,haselwandter2015}.
Unless indicated otherwise we use, consistent with experiments on glycine receptors and gephyrin scaffolds  \cite{specht2008,choquet2003,triller2008,triller2005,calamai2009,meier2001,haselwandter2011,haselwandter2015}, the diffusion coefficients $\nu_r=10^2\nu_s=10^{-2} \mu \text{m}^2/\text{s}$,
with the corresponding hopping rates $1/\tau_{\alpha}=2 \nu_{\alpha}/a^2$ in eq.~(\ref{eq:master}). The nonlinear diffusion terms in eqs.~(\ref{eq:mfield1}) and~(\ref{eq:mfield2}) result \cite{haselwandter2011,haselwandter2015} from crowding of distinct protein species. Mathematically equivalent terms arise in population biology \cite{satulovsky1996,mckane2004,lugo2008} and general models of non-Fickian diffusion \cite{fanelli2010,fanelli2013}. 
In line with experiments and large-scale computer simulations of crowded membranes \cite{metzler2016,jeon2016}, the nonlinear diffusion terms in eqs.~(\ref{eq:mfield1}) and~(\ref{eq:mfield2}) have been shown \cite{fanelli2010} to result in mean-square displacement curves bearing signatures of anomalous diffusion.

\begin{figure}[t!]
\onefigure[width=\columnwidth]{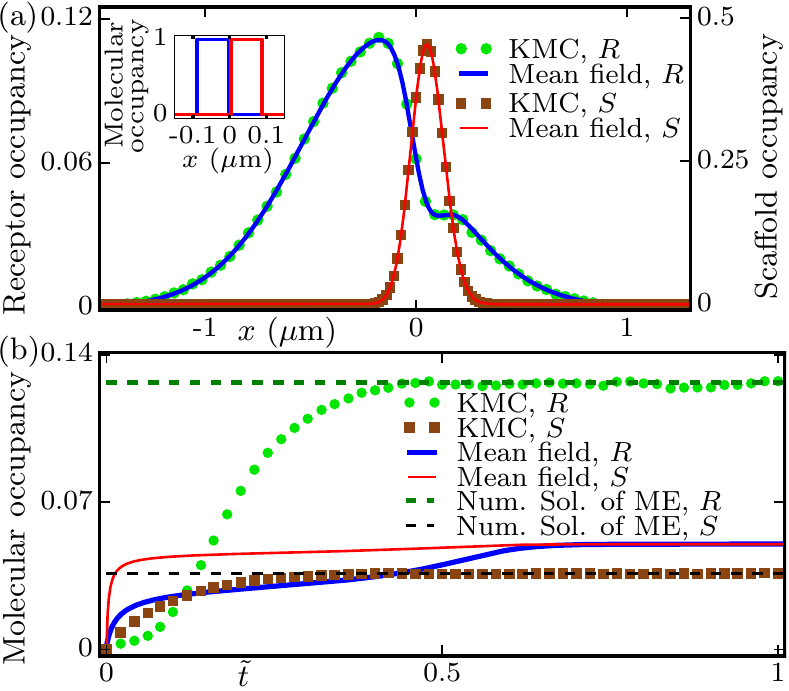}
\caption{Reaction and diffusion dynamics in crowded membranes. (a) Receptor and scaffold diffusion profiles at $t=10$~s using the initial conditions shown in the inset, obtained via KMC simulations of the ME in eq.~(\ref{eq:master}) with $W_\text{react}=0$ and the mean-field model in eqs.~(\ref{eq:mfield1}) and~(\ref{eq:mfield2}) with $F^r=F^s=0$ for $\nu_r=32 \, \nu_s=0.01$~$\mu\text{m}^2/\text{s}$. (b) Reaction dynamics of receptors and scaffolds \cite{haselwandter2011,haselwandter2015} in a single membrane patch obtained via KMC simulations and steady-state numerical solution of the ME in eq.~(\ref{eq:master}) with $W_\text{diff}=0$ and the mean-field model in eqs.~(\ref{eq:mfield1}) and~(\ref{eq:mfield2}) with $\nu_r=\nu_s=0$ \textit{vs.} scaled time $\tilde{t}=t/\tau$, with $\tau=1.3\times10^4$~s and $\tau=4.3\times10^5$~s for the stochastic and mean-field models, respectively. In (a,b), KMC simulations were averaged over $2\times10^3$ and $2\times10^4$ independent realizations, respectively.
}
\label{fig1}
\end{figure}

\section{Reaction and diffusion dynamics in crowded membranes}
Prior to studying the coupled reaction-diffusion dynamics at synaptic domains we consider here, in turn, reaction and diffusion processes in crowded membranes (see fig.~\ref{fig1}). We first consider the diffusion-only case corresponding
to eq.~(\ref{eq:master}) with $W_\text{react}=0$ and eqs.~(\ref{eq:mfield1}) and~(\ref{eq:mfield2}) with $F^r=F^s=0$ (see fig.~\ref{fig1}(a)). We find that the mean-field model of diffusion in crowded membranes in eqs.~(\ref{eq:mfield1}) and (\ref{eq:mfield2})
\cite{satulovsky1996,mckane2004,lugo2008,fanelli2010,fanelli2013,haselwandter2011,haselwandter2015}
produces non-Gaussian concentration profiles that are in excellent agreement with the corresponding average receptor and scaffold concentration profiles obtained from the stochastic lattice model in eq.~(\ref{eq:master}). Compared to Fickian diffusion, crowding yields less disperse molecule distributions, with scaffolds acting as an effective barrier to dispersal of the more rapidly diffusing receptors 
\cite{meier2001,borgdorff2002,dahan2003,hanus2006,triller2005,specht2008,triller2008,choquet2013,ziv2014,salvatico2015}.
In contrast, for the reaction-only case corresponding to eq.~(\ref{eq:master}) with $W_\text{diff}=0$ and eqs.~(\ref{eq:mfield1}) and~(\ref{eq:mfield2}) with $\nu_r=\nu_s=0$,
the mean-field model in eqs.~(\ref{eq:mfield1}) and (\ref{eq:mfield2}) fails to capture the temporal evolution as well as steady-state values of the average receptor and scaffold concentrations in eq.~(\ref{eq:master}) (see fig.~\ref{fig1}(b)), with the predicted mean-field dynamics being approximately
one order of magnitude slower than the average stochastic dynamics. 
We find that similar disagreement between the stochastic and mean-field
reaction dynamics persists even if $\epsilon$ is decreased to values substantially smaller than relevant for synaptic domains \cite{sm}. Consistent with previous studies \cite{samoilov2006,erban2009} demonstrating breakdown of mean-field models of cellular reaction dynamics, our results therefore suggest that molecular noise plays a central role in the reaction dynamics at synaptic domains.

\section{Collective fluctuations in synaptic domains}
\begin{figure}[t!]
\onefigure[width=\columnwidth]{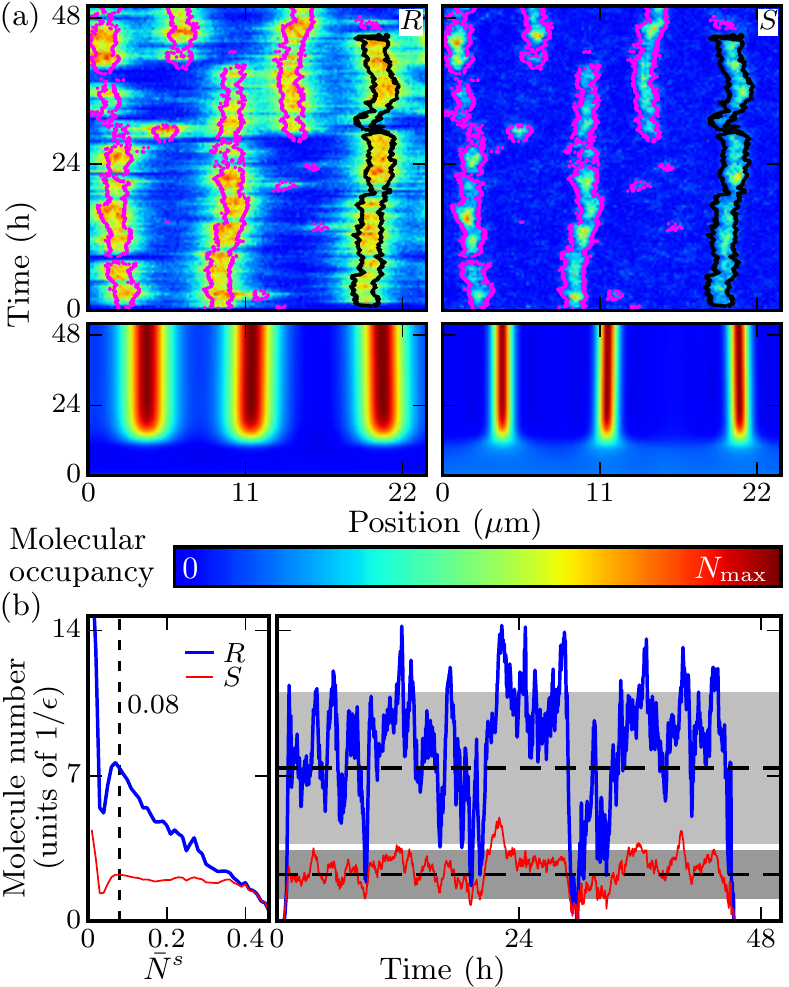}
\caption{Collective fluctuations in synaptic domains. (a) Synaptic domains obtained from KMC simulations of the ME in eq.~(\ref{eq:master}) with eqs.~(\ref{eq:diff}) and~(\ref{eq:react}) (upper panels) and the mean-field model in eqs.~(\ref{eq:mfield1}) and (\ref{eq:mfield2}) (lower panels) with maximum receptor and scaffold occupancies $(N^r_i,N^s_i)=(0.80, 0.57)$ (KMC) and $(r,s)=(0.46,0.15)$ (mean field). The curves in the upper panels show the domain boundaries obtained using a threshold $\bar N^s=0.08$ on the scaffold occupancy of membrane patches. (b) Average receptor and scaffold numbers per synaptic domain \textit{vs.} $\bar N^s$ (left panel) and receptor and scaffold numbers for the domain delineated by black domain boundaries in (a) \textit{vs.} time using $\bar N^s=0.08$ with the horizontal dashed lines and shaded areas indicating the average and standard deviation of receptor and scaffold numbers per domain (right panel). Averages and standard deviations in (b) were obtained from the domains in the upper panels of (a).}
\label{fig2}
\end{figure}

The stochastic lattice model of receptor and scaffold reaction-diffusion dynamics in eq.~(\ref{eq:master}) with eqs.~(\ref{eq:diff}) and~(\ref{eq:react})
and the corresponding mean-field model in eqs.~(\ref{eq:mfield1}) and (\ref{eq:mfield2}) both yield self-assembly of in-phase receptor and scaffold domains (see fig.~\ref{fig2}(a)) via \cite{haselwandter2011,haselwandter2015}
a reaction-diffusion (Turing) instability \cite{turing1952}. For the reaction and diffusion processes considered here, which in 2D can yield \cite{haselwandter2011,haselwandter2015} synaptic domains of the characteristic area found in experiments 
\cite{kirsch1995,meier2000,meier2001,borgdorff2002,dahan2003,hanus2006,ehrensperger2007,calamai2009,haselwandter2011}, eqs.~(\ref{eq:mfield1}) and~(\ref{eq:mfield2}) produce irregular but stable 1D patterns of synaptic domains with a characteristic wavelength $\approx 8.5$~$\mu$m, in line with the linear stability analysis of eqs.~(\ref{eq:mfield1}) and~(\ref{eq:mfield2}) \cite{haselwandter2011,haselwandter2015}. Equation~(\ref{eq:master}) yields a similar characteristic wavelength but produces substantial fluctuations in the size and location of synaptic domains, over a time scale $\approx10$~h.
Consistent with the results on the reaction-only system in fig.~\ref{fig1}(b), we find that molecular noise accelerates synaptic domain formation by approximately one order of magnitude compared to mean-field~dynamics \cite{sm}.

Scaffold concentration profiles across synaptic domains are less broad \cite{haselwandter2011,haselwandter2015}
than receptor concentration profiles (fig.~\ref{fig2}(a)). It is therefore convenient to define domain boundaries in terms of the scaffold density. We first apply a Savitzky-Golay filter \cite{savitzky1964} (order 5, frame size 25) to reduce small-scale fluctuations, and then detect domain boundaries using a threshold $\bar N^s$ on the scaffold occupancy of membrane patches. Increasing the value of $\bar N^s$ from zero, we obtain a local maximum of the average in-domain receptor and scaffold populations, as well as domain size, at around $\bar N^s=0.08$ (see fig.~\ref{fig2}(b)), and thus use $\bar N^s=0.08$ to specify domain boundaries (fig.~\ref{fig2}(a)). We find that, while the reaction-diffusion mechanism of synaptic domain formation \cite{haselwandter2011,haselwandter2015} yields characteristic average in-domain receptor and scaffold population numbers as well as domain sizes, the feedback between reaction-diffusion dynamics and molecular noise produces pronounced fluctuations in the in-domain receptor and scaffold population numbers (fig.~\ref{fig2}(b)), over a time scale of hours.

\begin{figure}[t!]
\onefigure[width=\columnwidth]{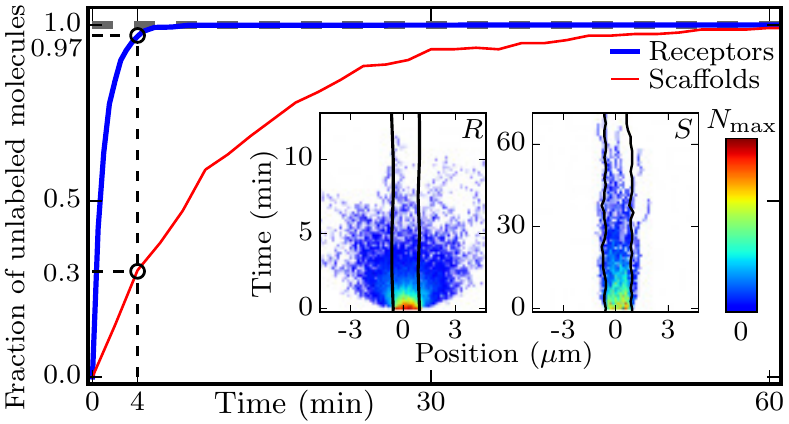}
\caption{Fraction of unlabeled receptors and scaffolds \textit{vs.} time for the representative synaptic domain delineated by black domain boundaries in fig.~\ref{fig2}(a) starting at $t\approx80$~min, at which time all the receptors and scaffolds
inside the synaptic domain are labeled. The insets show the membrane patch
occupancies of labeled receptors and scaffolds with maximum values $(N^r_i,N^s_i)=(0.59,0.21)$.
Black curves indicate domain boundaries, and membrane patches with vanishing numbers of labeled molecules are colored in white.
}
\label{fig3}
\end{figure}

\section{Molecular turnover}
Experiments have shown \cite{choquet2003,specht2008,triller2005,triller2008,ribrault2011,choquet2013,ziv2014,salvatico2015}
that synaptic domains are in a dynamic steady state, with individual receptors and scaffolds exchanging between synaptic domains and membrane regions outside
synaptic domains or the cytoplasm on typical time scales as short as seconds. To quantify molecular turnover in our stochastic lattice model \cite{haselwandter2011,haselwandter2015} we proceeded as in FRAP experiments \cite{choquet2013,calamai2009,specht2008}, and labeled all the receptors and scaffolds inside a synaptic domain at a given time. We then monitored the temporal evolution of the fraction of unlabeled receptors and scaffolds inside the synaptic domain (see fig.~\ref{fig3}). Adjusting the rate of receptor endocytosis in our model within the range of values suggested by experiments \cite{haselwandter2011, haselwandter2015,choquet2003,specht2008,triller2008} to $k_1=1/750$~$\text{s}^{-1}$ (table~\ref{tab1}) we find that, consistent with experiments \cite{choquet2003,specht2008, calamai2009}, $\approx30\%$ of scaffolds, but $>90\%$ of receptors, are replaced within $4$~min. Furthermore, our model yields a typical turnover time $\approx7$~min ($\approx54$~min) for receptor (scaffold) populations. In agreement with experiments \cite{choquet2003,specht2008,triller2005,triller2008,ribrault2011,choquet2013,ziv2014,salvatico2015}, we find that receptors initially localized in synaptic domains tend to leave synaptic domains via diffusion, while scaffolds typically stay localized within synaptic domains over their lifetime at the membrane (see fig.~\ref{fig3}~(insets)).

\begin{figure}[t!]
\onefigure[width=\columnwidth]{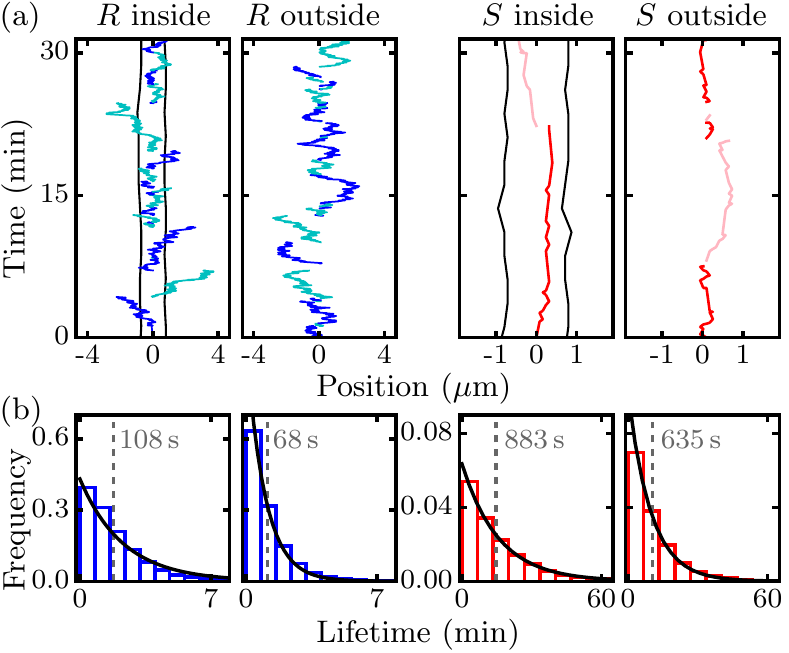}
\caption{Stochastic single-molecule dynamics at the membrane. (a) Sample membrane trajectories of receptors (light and dark blue curves) and scaffolds (pink and red curves) inserted inside and outside synaptic domains for the representative synaptic domain delineated by black domain boundaries in fig.~\ref{fig2}(a). Molecules are tracked at the membrane from insertion until removal. Black curves indicate domain boundaries. (b) Normalized frequency (in units of min$^{-1}$) of receptor and scaffold lifetimes at the membrane for molecules inserted inside synaptic domains close to the domain center and outside synaptic domains obtained from $10^4$ molecules each using KMC simulations of the ME in eq.~(\ref{eq:master}) with eqs.~(\ref{eq:diff}) and~(\ref{eq:react}). Average molecule lifetimes at the membrane are indicated by vertical dashed lines. The black solid curves show the exponential distributions of receptor and scaffold lifetimes at the membrane, $k^{r,s}_{\text{decay}} e^{-k^{r,s}_{\text{decay}} t_0}$, estimated theoretically using the approximation that the receptor and scaffold concentrations are uniform inside and outside synaptic domains (see the main text for further details).
}
\label{fig4}
\end{figure}

\section{Single-molecule dynamics}
To quantify single-molecule dynamics in our stochastic lattice model we follow the membrane trajectories of individual receptors and scaffolds inserted inside and outside synaptic domains (see fig.~\ref{fig4}(a)) and compute the distributions of lifetimes at the membrane for each molecule population (see fig.~\ref{fig4}(b)). Consistent with experiments 
\cite{triller2008,choquet2013,choquet2003,meier2001,borgdorff2002,dahan2003,hanus2006,specht2008,triller2005}
and the results in fig.~\ref{fig3}~(insets), we find that scaffolds inserted inside synaptic domains rarely leave the domain via diffusion, while receptors can readily diffuse into and out of synaptic domains. Receptors can diffuse over distances $\approx4$~$\mu\text{m}$ in our simulations, while scaffolds are typically confined to membrane regions $<1$~$\mu\text{m}$. We find that the average lifetime of receptors (scaffolds) inserted inside synaptic domains is approximately $60\%$ ($40\%$) longer than outside synaptic domains. Thus, the reaction-diffusion dynamics at synaptic domains can stabilize individual receptors and scaffolds at the membrane even if no stable molecular complexes are formed~\cite{ehrensperger2007, calamai2009, specht2013}.

The distributions of receptor and scaffold lifetimes in fig.~\ref{fig4}(b) can be understood by noting that, based on experiments on glycine receptors and gephyrin scaffolds, the stochastic lattice model used here \cite{haselwandter2011,haselwandter2015} allows for two key reactions removing receptors and scaffolds from the membrane. On the one hand, endocytosis of receptors and scaffolds yields an exponential decay of receptor and scaffold numbers at the membrane with rate constants $k_1$ and $k_6$, respectively (table~\ref{tab1}). On the other hand, receptors and scaffolds may be removed from the membrane, with rate constants $k_3$ and $k_8$ (table~\ref{tab1}), via binding of a bulk molecule or some other process that temporarily increases local crowding at the membrane. For membrane regions with an approximately homogeneous concentration of receptors and scaffolds, receptor and scaffold decay can then be approximated by a Poisson process with total effective rate $k^{r}_{\text{decay}}=k_1+k_3 \left(1-\left\langle N^r_i+N^s_i \right\rangle\right)$ for receptors and $k^{s}_{\text{decay}}=k_6+k_8 \left(1-\left\langle N^r_i+N^s_i \right\rangle\right)$ for scaffolds, where the average in-domain membrane patch occupancy $\left\langle N^r_i+N^s_i \right\rangle\approx0.62$ for the upper panels of fig.~\ref{fig2}(a) and
we use $\left\langle N^r_i+N^s_i \right\rangle\approx0$ outside synaptic domains. This simple argument yields exponential distributions of receptor and scaffold lifetimes at the membrane, $k^{r,s}_{\text{decay}} e^{-k^{r,s}_{\text{decay}}t_0}$,
where $t_0$ is the receptor or scaffold lifetime at the membrane, that match the histograms in fig.~\ref{fig4}(b) inside and outside synaptic domains, with no fitting parameters (fig.~\ref{fig4}(b)). The reaction and diffusion properties of synaptic receptors and scaffolds
\cite{kirsch1995,meier2000,meier2001,borgdorff2002,dahan2003,hanus2006,ehrensperger2007,calamai2009,haselwandter2011,haselwandter2015} thus provide a simple physical mechanism for spatially inhomogeneous receptor and scaffold lifetimes at the membrane, which is thought \cite{czondor2012,blanpied2002,earnshaw2006} to be a key feature of synaptic domains.

\section{Conclusion}
While the synaptic apparatus constitutes an enormously complex molecular machinery \cite{legendre2001,tyagarajan2014}, the reaction and diffusion properties of synaptic receptors and scaffolds are already sufficient 
\cite{kirsch1995,meier2000,meier2001,borgdorff2002,dahan2003,hanus2006,ehrensperger2007,calamai2009,haselwandter2011,haselwandter2015} for the self-assembly of synaptic receptor domains. In common with other
self-assembled biological structures, synaptic domains emerge in the presence of rapid stochastic turnover of their molecular components \cite{meier2001,borgdorff2002,dahan2003,hanus2006,triller2005,specht2008,triller2008}.
Based on the reaction and diffusion properties of synaptic receptors and scaffolds suggested by previous experiments and mean-field calculations
\cite{kirsch1995,meier2000,meier2001,borgdorff2002,dahan2003,hanus2006,ehrensperger2007,calamai2009,haselwandter2011,haselwandter2015}, we find here that the stochastic reaction-diffusion dynamics of receptors and scaffolds provide a simple physical mechanism for collective fluctuations in synaptic domains \cite{ribrault2011,choquet2013}, the molecular turnover observed at synaptic domains \cite{choquet2003,specht2008,calamai2009}, key features of the observed single-molecule trajectories \cite{choquet2003,choquet2013,meier2001,borgdorff2002,dahan2003,hanus2006,specht2008,triller2005,triller2008}, and spatially inhomogeneous receptor and scaffold lifetimes at the membrane \cite{czondor2012,blanpied2002,earnshaw2006}. While we focused here on the conceptually and computationally most straightforward scenario of a 1D system,
and showed that this 1D formulation already captures the basic phenomenology of the observed fluctuations at synaptic domains, a more detailed and quantitative model of the observed stochastic reaction-diffusion dynamics of synaptic receptors and scaffolds will necessitate a 2D formulation. Taken together, our results suggest that central aspects of the single-molecule and collective dynamics observed for membrane protein domains \cite{lang10,simons2010,rao2014,recouvreux2016,choquet2003} can be understood in terms of stochastic reaction-diffusion processes at the cell membrane, bringing us closer to a quantitative understanding of the organizational principles linking the collective properties of biologically important supramolecular structures to the stochastic dynamics that rule their molecular components.

\acknowledgments
We thank J. Q. Boedicker, R. A. da Silveira, M. Kardar, and A. Triller for helpful discussions. This work was supported by NSF award numbers DMR-1554716 and DMR-1206332, an Alfred P. Sloan Research Fellowship in Physics, the James H. Zumberge Faculty Research and Innovation Fund at USC, and the USC Center for High-Performance Computing. We also acknowledge support through the Kavli Institute for Theoretical Physics, Santa Barbara, via NSF award number PHY-1125915.




\section*{Supplementary material}

# Stochastic single-molecule dynamics of synaptic membrane protein domains
## Supplementary Material

Osman Kahraman, Yiwei Li, and Christoph A. Haselwandter

*Department of Physics & Astronomy and Molecular and Computational Biology Program,*
*Department of Biological Sciences, University of Southern California, Los Angeles, CA 90089, USA*

## STOCHASTIC DYNAMICS OF SYNAPTIC DOMAINS IN 2D

To check that our stochastic lattice model gives similar fluctuations at synaptic domains in 2D as in 1D, we repeated our KMC simulations for the 2D version of the reaction-diffusion dynamics described in the main text [1, 2]. We find that receptors and scaffolds self-assemble into domains with the size and molecule occupancies expected based on our 1D simulations (see fig. S1). Importantly, receptor-scaffold domains are still subject to substantial fluctuations in size and location in 2D, over spatial and temporal scales consistent with our 1D simulations.

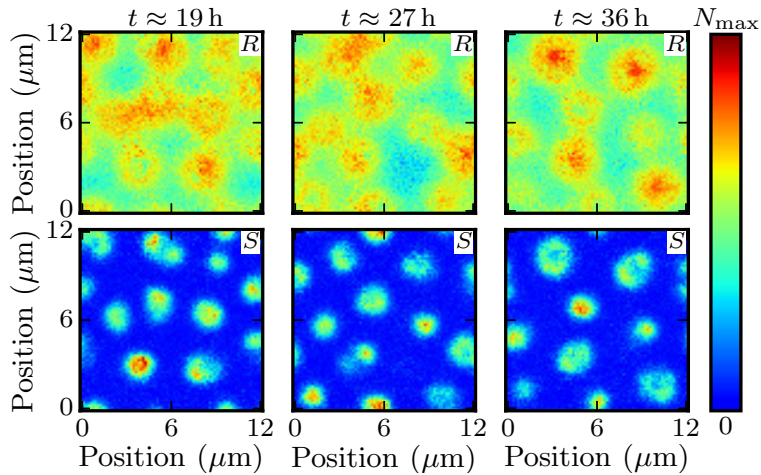

FIG. S1: Snapshots from a 2D KMC simulation using the same reaction and diffusion processes and parameter values as in the main text. Similarly as in the main text, we employed periodic boundary conditions and random initial conditions. We have the maximum receptor and scaffold occupancies $\left(N_{i,j}^r, N_{i,j}^s\right) = (0.82, 0.62)$, where the subscripts $(i,j)$ denote the lattice sites in the 2D (square) lattice used here.

# EFFECT OF PROTEIN NUMBER ON STOCHASTIC DYNAMICS

We repeated the KMC simulations in figs. 1(b) and 2 of the main text for different values of the normalization constant $\epsilon$, to determine how our stochastic lattice model compares with the corresponding mean-field model as the maximum number of proteins per membrane patch is increased, i.e., $\epsilon$ is decreased. As discussed in the main text, the inverse of $\epsilon$ sets the maximum number of molecules that a membrane patch can accommodate. Note that the limit $\epsilon \to 0$ is expected to be of limited physical relevance for synaptic domains (as well as other membrane protein domains), because the patch size must necessarily be smaller than the expected size of synaptic domains which, together with the approximate size of glycine receptors and gephyrin, sets a lower limit on the value of $\epsilon$, i.e., an upper limit on the number of receptor or scaffold molecules a membrane patch can accommodate. As discussed in the main text we set, for the calculations described in the main text, the values of $\epsilon$ and $a$ to lie within the regime that is relevant for synaptic domains, based on the approximate size of glycine receptors and gephyrin.

In analogy to fig. 1(b) of the main text, fig. S2 compares averages over KMC simulations with the corresponding mean-field results for the reaction dynamics at synaptic domains, using $\epsilon = 1/300$ and $\epsilon = 1/600$ as well as the value $\epsilon = 1/100$ corresponding to fig. 1(b) of the main text. We find that, even for $\epsilon = 1/600$, the mean-field model fails to capture the steady-state as well as the temporal evolution of the stochastic model, with the predicted mean-field dynamics being approximately one order of magnitude slower than the average stochastic dynamics. We also carried

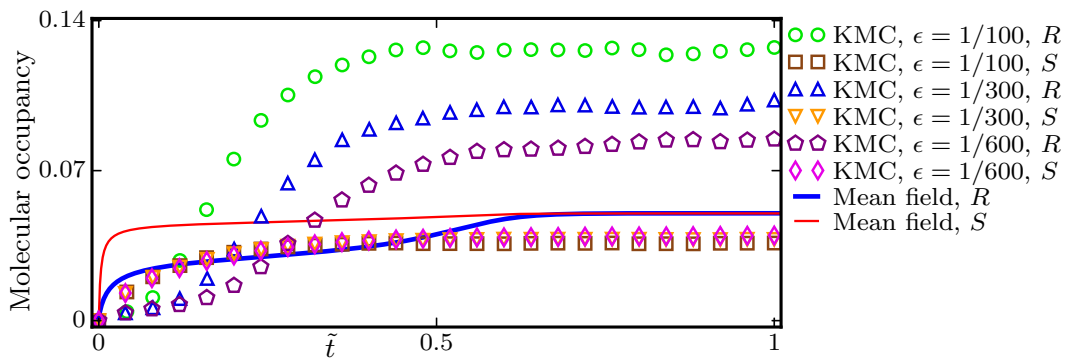

FIG. S2: Reaction dynamics of receptors and scaffolds in a single membrane patch obtained, as in fig. 1(b) of the main text, via KMC simulations of the ME in eq. (1) of the main text with $W_{\text{diff}} = 0$ using $\epsilon = 1/100$, $\epsilon = 1/300$, and $\epsilon = 1/600$, and the mean-field model in eqs. (4) and (5) of the main text with $\nu_r = \nu_s = 0$ vs. scaled time $\tilde{t} = t/\tau$, with $\tau = 1.3 \times 10^4$ s and $\tau = 4.3 \times 10^5$ s for the stochastic and mean-field models, respectively. The KMC simulations were averaged over $2 \times 10^4$ independent realizations.

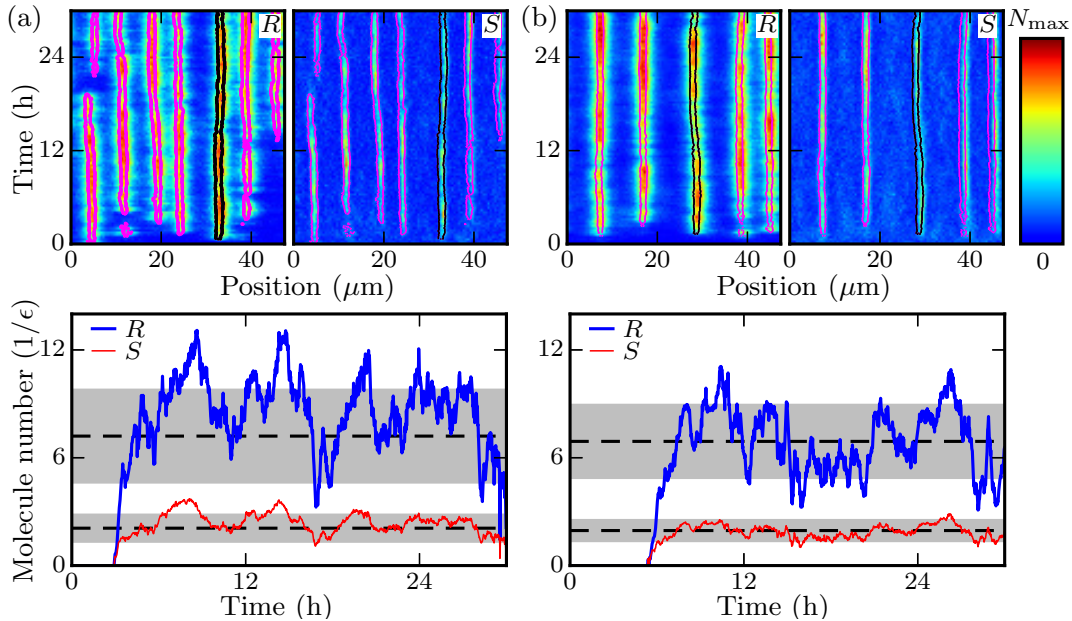

FIG. S3: KMC simulations as in fig. 2 of the main text, but using (a) $\epsilon = 1/300$ and (b) $\epsilon = 1/600$. The upper panels show the synaptic domains obtained from KMC simulations, and the lower panels show the receptor and scaffold numbers for the domains delineated by black domain boundaries in the upper panels *vs.* time using a threshold $\bar{N}_s = 0.08$ on the scaffold occupancy of membrane patches. The horizontal dashed lines and shaded areas in the lower panels indicate the average and standard deviation of receptor and scaffold numbers per domain, obtained from the domains in the upper panels. In the upper panels we have the maximum molecule occupancies (a) $(N_i^r, N_i^s) = (0.69, 0.42)$ and (b) $(N_i^r, N_i^s) = (0.64, 0.32)$.

out KMC simulations of the reaction-diffusion dynamics for decreased values of $\epsilon$. Again, in addition to the value $\epsilon = 1/100$ used for fig. 2 of the main text, we consider $\epsilon = 1/300$ and $\epsilon = 1/600$ (see fig. S3). Although we find that fluctuations in the size and location of synaptic domains decrease somewhat with decreasing $\epsilon$ (fig. S3 (upper panels)), we still obtain substantial fluctuations in the in-domain receptor and scaffold populations (fig. S3 (lower panels)). Furthermore, we find that the timescale of domain formation in the stochastic system is comparable to our results in fig. 2(a) of the main text, and is consistent with the more rapid progression of the stochastic reaction dynamics, compared to the corresponding mean-field results, in fig. S2.



\end{document}